\documentclass[doublecol]{epl2} 

\usepackage{amsmath}
\usepackage{mathtools}
\usepackage{amsfonts}
\usepackage{amssymb}
\usepackage{graphicx}
\usepackage{lmodern}
\usepackage{cite}
\usepackage{csquotes}
\usepackage{hyperref}
\usepackage{dcolumn}
\usepackage{booktabs}
\usepackage{bm}
\usepackage[mathlines]{lineno}
\usepackage{float}
\usepackage[T1]{fontenc}
\usepackage{subfigure}
\title{A Rotational Perturbative Correction to Democratic Neutrino Mixing and JUNO Compatibility }

\author{  Maibam Ricky Devi\inst{1} \and Swaraj Kumar Nanda\inst{2} \and Sudhanwa Patra\inst{3}}

\institute{                    
  \inst{1}  Department of Physics, Gauhati University,  Guwahati-781014, Assam, India\\
   \inst{2} Department of Physics, ITER, SOA University, Bhubaneswar-751030, India, \\
    \inst{3} Department of Physics, Indian Institute of Technology Bhilai, Durg-491002, India
}

\abstract{In this work, we revamp the democratic mixing matrix (DM) by adding a perturbation term such that the mixing angles derived from it are compatible with the NuFIT 6.1 and recent findings from the Jiangmen Underground Neutrino Observatory (JUNO). To do this, we have incorporated perturbation term in the elements of the mixing matrix such that it does not lose its unitarity. Thus, the democratic mixing matrix, once ruled out by the experimental evidences from T2K, Double Chooz, and Daya Bay, can be modified with rotational perturbation in the (1,2), (1,3), and (2,3) sectors and additional real parameters added in each element of DM. Finally, we analyze the allowed and disallowed textures in light of the JUNO findings.}
\begin{document}
\maketitle

\newcolumntype{P}[1]{>{\centering\arraybackslash}p{#1}}

\section{Introduction}
In 1988, Harald Fritzsch proposed in a conference \cite{Fritzsch:1988ix} that both quarks and charged leptons can be described by a mass matrix in which all elements have equal mass 
\begin{equation}
    M= \dfrac{m}{3} \left(\begin{matrix}
        1 & 1 & 1 \\
          1 & 1 & 1 \\
            1 & 1 & 1 
    \end{matrix}\right)
    \label{eqn1}
\end{equation}
along with some additional terms.
This matrix can be diagonalized with the help of a mixing matrix given by
\begin{equation}
    U= \left( \begin{matrix}
       \dfrac{1}{\sqrt{2}} & -\dfrac{1}{\sqrt{2}}  & 0\\
       \dfrac{1}{\sqrt{6}} & \dfrac{1}{\sqrt{6}} & -\dfrac{2}{\sqrt{6}} \\
       \dfrac{1}{\sqrt{3}} &  \dfrac{1}{\sqrt{3}} &  \dfrac{1}{\sqrt{3}} 
    \end{matrix}\right)
\end{equation}
The matrix of the type in Eqn. \ref{eqn1} was named \enquote{democratic} by C. Jarlskog \cite{Jarlskog:1987vh}. Later in 1996,  Harald Fritzsch and Zhi-zhong Xing  suggested that the same mixing matrix $U$ can be used in the neutrino sector to explain its tiny mass generation and flavor mixing \cite{Fritzsch:1995dj}. The $ 3\times 3$ lepton flavor mixing pattern is obtained as the leading term after the democratic flavor symmetry breaking or $S(3)_L \times S(3)_R$ of the charged-lepton mass matrix  provided the Majorana mass matrix is diagonal. This ansatz predicted three nearly degenerate neutrino masses and large mixing angles, $ \theta_{12}={45}^\circ, \;  \theta_{23}={54.7}^\circ$, which are ruled out by the present neutrino oscillation data. However, if proper correction is provided using perturbation, it can be made compatible with the latest NuFIT 6.1 data \cite{NuFIT6.1}.  \\
Several work based on the democratic-mixing ansatz have been done, including the incorporation of democracy mixing within a universal seesaw framework by Koide \cite{Koide:1996me, Koide:1997iw}, with an attempt to explain a common origin for quark and lepton mixing. Some interesting work based on democratic mixing can be found in the literature \cite{Fukuura:1999ze, Koide:2000fg,Miura:2000bj,  Joaquim:2000fx}.
\\
During 2002-2005, several studies shifted their focus to determine a more realistic mixing from partial democracy \cite{Koide:2002nc}, symmetry underlying the neutrino mixing \cite{Harrison:2003he}, third-generation dominance scenario \cite{Dermisek:2003rw} and from radiative and seesaw-threshold corrections \cite{Mei:2005gp}.
\\
The experimental measurement of the neutrino mixing angles opened the possibility to reconcile the democratic structure using different approaches of perturbation method such as Xing \cite{Xing:2010pn, Xing:2011at} attempted to accommodate the democratic structure in TBM (tri-bimaximal mixing) like structure using natural perturbation, while Deepthi and Mohanta \cite{Deepthi:2012zt} studied a generalized perturbation in DM (democratic mixing) to observe possible leptonic CP violation.

A rotation based correction were carried out by Garg and Gupta \cite{Garg:2013xwa} where the parametrization induced by rotation $R_{ij}.U.R_{kl}$ can give viable perturbative structures. Here, $R_{ij},\; R_{kl}$ are rotations in $i-j$ and $k-l$ sectors while $U$ is any special mixing matrix. Later in 2018, Garg in his work \cite{Garg:2017mjk} included  double-rotation corrections to test the perturbed democratic structure against the global-fit values of neutrino oscillation parameters. 

In our work.  we incorporate a similar rotation-based correction in the democratic mixing structure along with an additional perturbed term in the unperturbed democratic mixing matrix. This is done such that the democratic structure does not lose it unitarity.  We implement the rotation in both the unperturbed and perturbed democratic mixing structure and test the viable structures in the next two sections.
\\
We outline our paper as follows. In section 2, we discuss the viability of textures for rotation in 1-2, 1-3 and 2-3 sectors in unperturbed democratic mixing. In section 3, we discuss the viable rotationally perturbed democratic textures against the NuFIT 6.1 \cite{NuFIT6.1} data. In section 4, we will present our numerical analysis and result, and finally in section 5, we will present our summary and conclusion of our work. 

\section{Testing viability of rotation in unperturbed democratic mixing}
We define the undefined democratic mixing as $U_{DM}$ and subject it to rotation along the sectors (1-2), (1-3) and (2-3) similar to TM1 and TM2 structures \cite{Harrison:2002er, Xing:2002sw, He:2003rm}. We define them as $V_a, \; V_b \; \text{and} \; V_c$ respectively. 
\begin{equation}
    U_{DM}= \left(
\begin{array}{ccc}
 \frac{1}{\sqrt{2}} & -\frac{1}{\sqrt{2}} & 0 \\
 \frac{1}{\sqrt{6}} & \frac{1}{\sqrt{6}} & -\sqrt{\frac{2}{3}} \\
 \frac{1}{\sqrt{3}} & \frac{1}{\sqrt{3}} & \frac{1}{\sqrt{3}} \\
\end{array}
\right)
\end{equation}
\begin{equation}
    R_{12}=\left(
\begin{array}{ccc}
 c  & e^{i \alpha } s & 0 \\
 -e^{-i \alpha } s & c & 0 \\
 0 & 0 & 1 \\
\end{array}
\right)
\end{equation}

\begin{eqnarray}
V_a &=& U_{DM} R_{12} \\ \nonumber
 V_a &=&  \left(
\begin{array}{ccc}
 \frac{c}{\sqrt{2}}+\frac{e^{-i \alpha } s}{\sqrt{2}} & \frac{e^{i \alpha } s}{\sqrt{2}}-\frac{c}{\sqrt{2}} & 0 \\
 \frac{c}{\sqrt{6}}-\frac{e^{-i \alpha } s}{\sqrt{6}} & \frac{c}{\sqrt{6}}+\frac{e^{i \alpha } s}{\sqrt{6}} & -\sqrt{\frac{2}{3}} \\
 \frac{c}{\sqrt{3}}-\frac{e^{-i \alpha } s}{\sqrt{3}} & \frac{c}{\sqrt{3}}+\frac{e^{i \alpha } s}{\sqrt{3}} & \frac{1}{\sqrt{3}} \\
\end{array}
\right)
\end{eqnarray}
   where, $ c=\cos\theta , \; \text{and} \; s=\sin \theta $ . \\
   \\
The $V_a$ mixing is not viable because here $V_{e3}=0$.\\
\\
\begin{equation}
  R_{13}= \left(
\begin{array}{ccc}
 c & 0 & e^{i \alpha } s \\
 0 & 1 & 0 \\
 -e^{-i \alpha } s & 0 & c \\
\end{array}
\right)
\end{equation}
\begin{eqnarray}
V_b &=& U_{DM} R_{13} \\ \nonumber
 V_b &=&  \left(
\begin{array}{ccc}
 \frac{c}{\sqrt{2}} & -\frac{1}{\sqrt{2}} & \frac{e^{i \alpha } s}{\sqrt{2}} \\
 \frac{c}{\sqrt{6}}+e^{-i \alpha } s \sqrt{\frac{2}{3}} & \frac{1}{\sqrt{6}} & \frac{e^{i \alpha } s}{\sqrt{6}}-\sqrt{\frac{2}{3}} c \\
 \frac{c}{\sqrt{3}}-\frac{e^{-i \alpha } s}{\sqrt{3}} & \frac{1}{\sqrt{3}} & \frac{c}{\sqrt{3}}+\frac{e^{i \alpha } s}{\sqrt{3}} \\
\end{array}
\right)
\end{eqnarray}
On Comparing $V_b$ with the standard $U_{PMNS}$ matrix, the $V_b$ mixing gives the mixing angle relations as $\sin^2 \theta_{13} =s^2/2$,  $\sin^2 \theta_{12} = \dfrac{1}{2(1-\sin^2\theta_{13})}$ and $\sin^2 \theta_{23} = \dfrac{\dfrac{2}{3} -\dfrac{1}{3}s^2 -\dfrac{2}{3}cs \cos \alpha}{1-\dfrac{1}{2}s^2}$. If we $\sin^2\theta_{13} \approx 0.0222 \in (0.0205-0.0246) \; NuFIT\; 3\sigma \; range$ then $\sin^2\theta_{12} \approx 0.511$. Since JUNO \cite{JUNO:2025gmd} gives $\sin^2 \theta_{12} = 0.309 \pm 0.0087$, thus the predicted solar angle from this mixing pattern is far above the $3\sigma$ interval (0.0205-0.0246).  Hence $V_b$ mixing pattern is being ruled out by the current NuFIT 6.1 oscillation data.
\\
\\
The mixing $V_c =U_{DM} R_{23}$ exhibits a texture as 
\begin{eqnarray}
  V_c &=& U_{DM} R_{23} \\ \nonumber
 V_c &=& \left(
\begin{array}{ccc}
 \frac{1}{\sqrt{2}} & -\frac{c}{\sqrt{2}} & -\frac{e^{i \alpha } s}{\sqrt{2}} \\
 \frac{1}{\sqrt{6}} & \frac{c}{\sqrt{6}}+e^{-i \alpha } s \sqrt{\frac{2}{3}} & \frac{e^{i \alpha } s}{\sqrt{6}}-\sqrt{\frac{2}{3}} c \\
 \frac{1}{\sqrt{3}} & \frac{c}{\sqrt{3}}-\frac{e^{-i \alpha } s}{\sqrt{3}} & \frac{c}{\sqrt{3}}+\frac{e^{i \alpha } s}{\sqrt{3}} \\
\end{array}
\right)
\end{eqnarray}
where, \begin{equation}
  R_{23}= \left(
\begin{array}{ccc}
 1 & 0 & 0 \\
 0 & c & e^{i \alpha } s \\
 0 & -e^{-i \alpha } s & c \\
\end{array}
\right)
\end{equation}
The solar mixing angle and the $\delta_{cp}$ phase in this type of mixing can be evaluated by comparing $V_c$ with the $U_{PMNS}$ matrix. Thus, we get
\begin{eqnarray}
\sin^2\theta_{13}&=&\dfrac{s^2}{2}
\\
    \sin^2\theta_{12} &=& \dfrac{1-2\sin^2\theta_{13}}{2(1-\sin^2\theta_{13})}   \label{eqn8-9}
    \\
     \sin^2\theta_{23} &=& \dfrac{\dfrac{2}{3}-\dfrac{1}{2}s^2 -\dfrac{2}{3}cs \cos \alpha}{1-\dfrac{1}{2}s^2} 
    \\
    \sin \delta_{cp} &=& \dfrac{cs \sin \alpha}{6\sqrt{3} c_{12} c_{23} c^2_{13} s_{12}s_{23}s_{13}}
    \end{eqnarray}

    Using the current best-fit value $\sin^2 \theta_{13} \approx 0.0222$ the $V_c$ mixing matrix predicts the solar mixing angle from Eqn. \ref{eqn8-9} as, $\sin^2\theta_{12} \approx 0.4886$ which lies outside the allowed NuFIT 6.1 interval. Hence this mixing matrix is excluded too by the current neutrino oscillation data.
    
\section{Rotational Corrections in the Perturbed Democratic Neutrino Mixing}
To fix the problems in the above mixing patterns, we implement perturbative modifications in the three mixing patterns such that the solar mixing angles obtained from them fall within the $3\sigma$ region of the new JUNO result. We follow the reparametrization technique originally introduced by Yuta Hyodo and Teruyuki Kitabayashi in their literature \cite{Hyodo:2023sku}. In this process, we introduce a new parameter $\epsilon$ in the mixing matrix $U_{DM}$ so that it does not lose its unitarity property, $U^{\dagger}_{DM}\;U_{DM}=1 \;  or \;U^{T}_{DM}\;U_{DM}=1 $ if $\epsilon$ is real. We represent the new modified democratic mixing matrix as $U_{MDM}$. We initiate the modification with the $U_{e2}$ element because our goal is to fit the solar mixing angle within the allowed range of JUNO data i.e., 
\begin{equation}
    -\dfrac{1}{\sqrt{2}} \; \rightarrow \; -\sqrt{\dfrac{1}{2} + \epsilon}
\end{equation}
Thus  the modified democratic mixing matrix $U_{MDM}$ satisfying the unitarity condition, can be written as:
\begin{equation}
   U_{MDM} =  \left(
\begin{array}{ccc}
 \sqrt{x_1+\frac{1}{2}} & -\sqrt{\epsilon +\frac{1}{2}} & 0 \\
 \sqrt{x_2+\frac{1}{6}} & \sqrt{x_3+\frac{1}{6}} & -\sqrt{x_4+\frac{2}{3}} \\
 \sqrt{x_5+\frac{1}{3}} & \sqrt{x_6+\frac{1}{3}} & \sqrt{x_7+\frac{1}{3}} \\
\end{array}
\right)
\end{equation}
This gives us a set of simultaneous equations whose solutions can yield the values of $x_i \; (i=1,2,...,7)$. This set of equations are as follows:
\begin{eqnarray}
 \nonumber
    x_1+x_2+x_5+1&=& 1\\ \nonumber
    x_3+x_6+\epsilon +1 &=& 1\\  \nonumber
    x_4+x_7+1 &=& 1\\  \nonumber
  \sqrt{x_1+\frac{1}{2}} \left(-\sqrt{\epsilon +\frac{1}{2}}\right)+  \sqrt{x_2+\frac{1}{6}} \sqrt{x_3+\frac{1}{6}}+  & & \\   \nonumber
 \sqrt{x_5+\frac{1}{3}} \sqrt{x_6+\frac{1}{3}} &=& 0  \\ \nonumber
 \sqrt{x_5+\frac{1}{3}} \sqrt{x_7+\frac{1}{3}}-\sqrt{x_2+\frac{1}{6}} \sqrt{x_4+\frac{2}{3}} &=& 0\\ \nonumber
 \sqrt{x_6+\frac{1}{3}} \sqrt{x_7+\frac{1}{3}}-\sqrt{x_3+\frac{1}{6}} \sqrt{x_4+\frac{2}{3}} &= & 0 
\end{eqnarray}
For our convenience, we take $x_4 =x_7=0$. Thus we get the values of the  $x_i$s in terms of $\epsilon$ as:
\begin{equation}
    (x_1,x_2,x_3,x_4,x_5,x_6,x_7 ) = (-\epsilon, \dfrac{\epsilon}{3},-\dfrac{\epsilon}{3},0, \dfrac{2\epsilon}{3}, -\dfrac{2\epsilon}{3},0)
\end{equation}
Thus the modified $U_{DM}$ can be written as
\begin{equation}
    U_{MDM}=\left(
\begin{array}{ccc}
 \sqrt{\frac{1}{2}-\epsilon } & -\sqrt{\epsilon +\frac{1}{2}} & 0 \\
 \sqrt{\frac{\epsilon }{3}+\frac{1}{6}} & \sqrt{\frac{1}{6}-\frac{\epsilon }{3}} & -\sqrt{\frac{2}{3}} \\
 \sqrt{\frac{2 \epsilon }{3}+\frac{1}{3}} & \sqrt{\frac{1}{3}-\frac{2 \epsilon }{3}} & \frac{1}{\sqrt{3}} \\
\end{array}
\right)
\end{equation}
\\
\\
Before we move on to implement rotational correction in the above modified democratic mixing angle, we would like to mention that rotational correction along the (1-2) sector in $U_{MDM}$, i.e., $ V^{\prime}_a=U_{MDM}.R_{12}$,  gives $U_{e3}=0$ which is not viable with the current experimental value. Hence, we proceed with the other two rotational corrections in the (1-3) and (2-3) sectors respectively. 

\subsection{Rotation in the (1-3) sector of the perturbed democratic structure: First Form}

Here we implement rotational correction into the perturbed democratic structure as :
\begin{equation}
    V^{\prime}_b=U_{MDM}.R_{13}
    \label{V2}
\end{equation}

\begin{equation}
   V^{\prime}_b = \left(
\begin{array}{ccc}
 A c & B & A e^{i \alpha } s \\
 -c G-\frac{e^{-i \alpha } s}{\sqrt{2}} & F & \frac{c}{\sqrt{2}}-e^{i \alpha } G s \\
 \frac{e^{-i \alpha } s}{\sqrt{2}}-c G & F & -\frac{c}{\sqrt{2}}-e^{i \alpha } G s \\
\end{array}
\right)
\label{Vb}
\end{equation}
\\
The standard matrix $U_{PMNS}$ for Majorana type neutrinos can be written as
\begin{eqnarray}
U_{PMNS} &=& \begin{bmatrix}
1 & 0 & 0\\
0 & c_{23} & s_{23}\\
0 & - s_{23} & c_{23}
\end{bmatrix}\times \begin{bmatrix}
c_{13} & 0 & s_{13}\,e^{-i\delta}\\
0 & 1 & 0\\
-s_{13} e^{i\delta} & 0 & c_{13}
\end{bmatrix}\nonumber\\
&& \quad\quad\times\begin{bmatrix}
c_{12} & s_{12} & 0\\
-s_{12} & c_{12} & 0\\
0 & 0 & 1
\end{bmatrix}.P,
\end{eqnarray}
where $P= diag(1,e^{i\alpha/2}, e^{i\beta/2})$.
Comparing the matrix $U_{PMNS}$ with Eqn. \ref{Vb}, we get
\begin{equation}
\boxed{
s_{13}^2=
\left(\frac12-\epsilon\right)\sin^2\theta
},
\label{eqn:theta}
\end{equation}
and
\begin{equation}
\boxed{
s_{12}^2=
\frac{\frac12+\epsilon}
{1-s_{13}^2}
}.
\label{s12:first}
\end{equation}

The perturbation parameter $\epsilon$ can be evaluated as
\begin{equation}
\boxed{
\epsilon=
s_{12}^2(1-s_{13}^2)-\frac12
}.
\label{eqn:epsilon}
\end{equation}

Finally the atmospheric mixing angle can be evaluated as
\begin{align}
s_{23}^2={1\over1-s_{13}^2}
\bigg[
&\left(\frac{\epsilon}{3}+\frac16\right)\sin^2\theta
+\frac23\cos^2\theta \nonumber\\
&-\frac23\sqrt{1+2\epsilon}\,
\sin\theta\cos\theta\cos\alpha
\bigg].
\label{eqn:theta23}
\end{align}

It is well known that the leptonic Jarlskog invariant is given by
\begin{equation}
J_{\rm CP}
=
\operatorname{Im}
\left[
U_{e1}U_{\mu2}U_{e2}^{*}U_{\mu1}^{*}
\right] .
\label{Jarlskog}
\end{equation}

In terms of standard parameterization, it can be written as
\begin{equation}
\boxed{
J_{\rm CP}
=
c_{12}c_{23}c_{13}^{2}
s_{12}s_{23}s_{13}\sin\delta
= J_{{CP}_{max}}\sin \delta}.
\label{eqn:J}
\end{equation}

The Jarlskog invariant obtained from the first form is 
\begin{equation}
\boxed{
J_{\rm CP}
=
-\frac{(1-2\epsilon)\sqrt{1+2\epsilon}}
{6\sqrt3}
\sin\theta\cos\theta\sin\alpha
}.
\label{eqn:J1}
\end{equation}

Using benchmark values of the mixing angles from NuFIT 6.1 as
\begin{equation}
s_{12}^2\simeq0.304,\qquad
s_{13}^2\simeq0.0223,\qquad
s_{23}^2\simeq0.573,
\end{equation}
we obtain the unknown parameters $\epsilon$ and $\theta$ from Eqns \ref{eqn:theta} and \ref{eqn:epsilon} as
\begin{equation}
\epsilon\simeq-0.203,
\end{equation}
and
\begin{equation}
\sin^2\theta
=
\frac{0.0223}{0.5-\epsilon}
\simeq 0.0317,
\end{equation}
this gives
\begin{equation}
\theta\simeq10.2^\circ.
\end{equation}

The atmospheric constraint of Eqn. \ref{eqn:theta23} leads to 
\begin{equation}
\cos\alpha\simeq 0.9815 \; \; \implies \; 
\alpha\simeq 11.02^\circ \; \text{or} \; 348.98^\circ.
\end{equation}

Thus, substituting these benchmark values of the parameters in Eqn. \ref{Jarlskog}, we get
\begin{equation}
J_{\rm CP}\simeq \; -0.0183 \sin\alpha ,
\end{equation}
and therefore, for the bound  $-0.1911 \; \leq \; \sin \alpha \; \leq \; 0.1911$, we  get
\begin{equation}
\boxed{
-0.0035\lesssim J_{\rm CP}\lesssim 0.0035
}.
\end{equation}
Thus, our first form of rotational perturbation in the modified democratic mixing can easily fit $J_{CP}$ within the allowed experimental range given by NuFIT 6.1.
\\
\begin{figure*}
  \centering
    \subfigure[]{\includegraphics[width=0.38\textwidth]{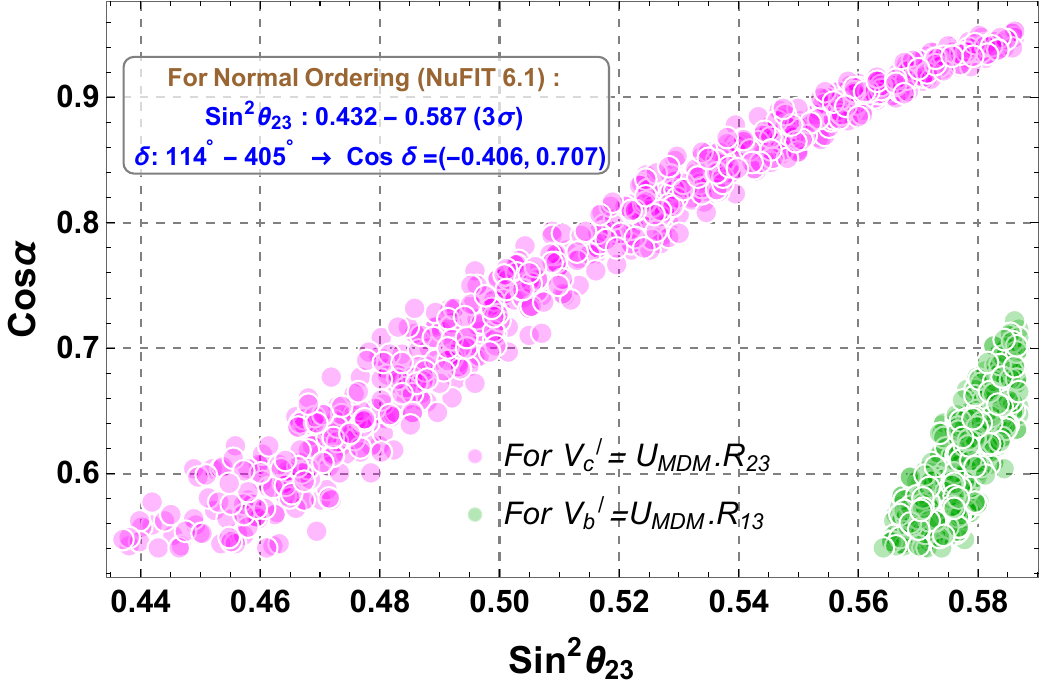}\label{fig:a}} 
    \subfigure[]{\includegraphics[width=0.38\textwidth]{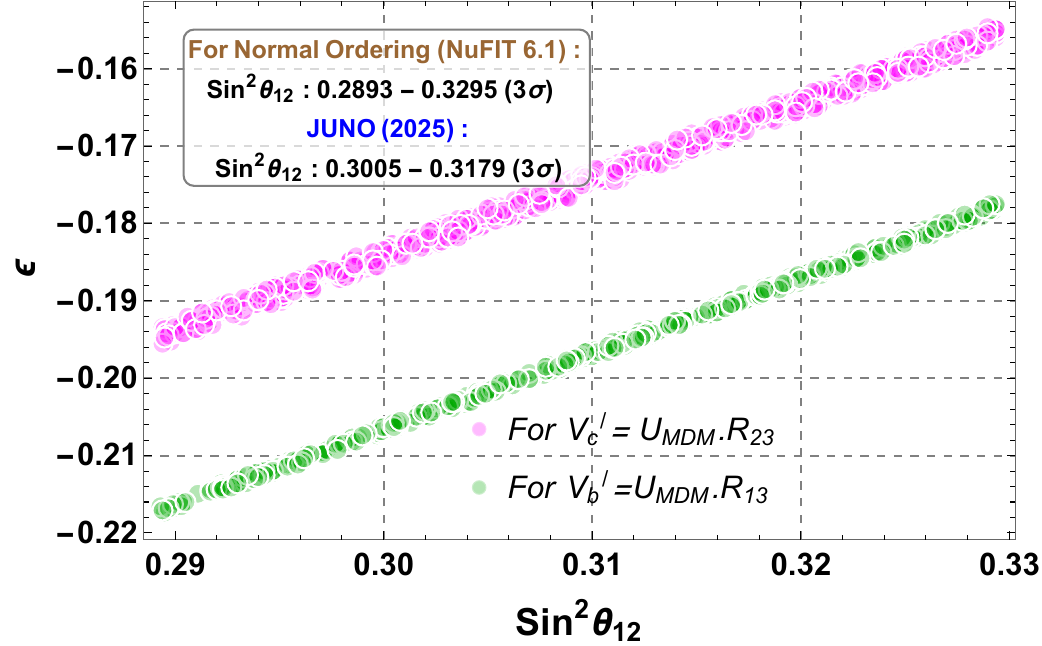}\label{fig:b}} 
    \subfigure[]{\includegraphics[width=0.38\textwidth]{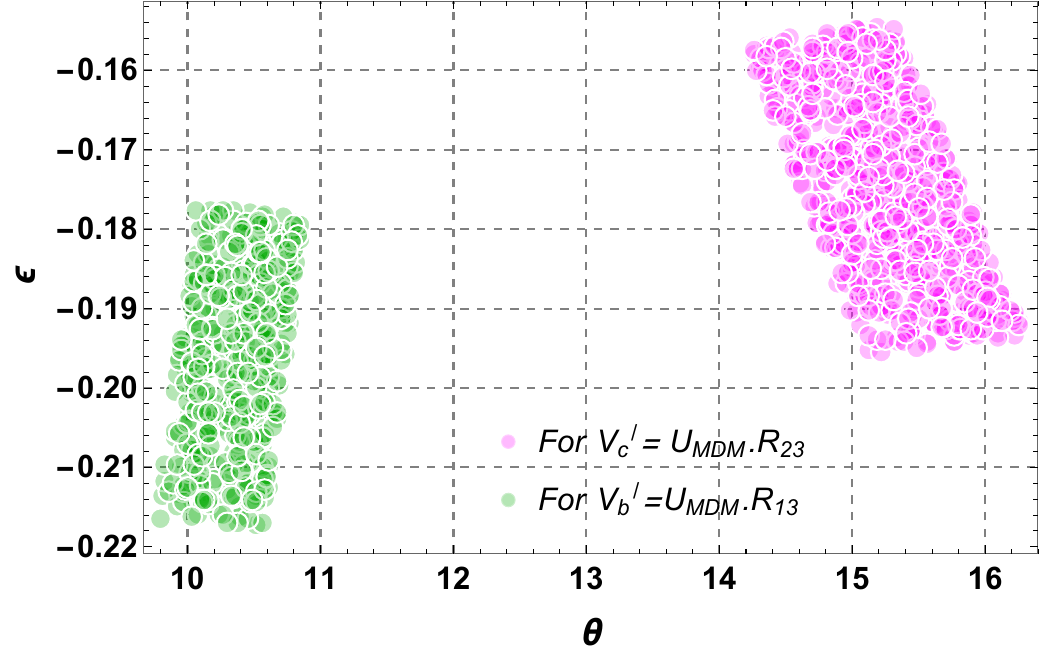}\label{fig:c}} 
    \subfigure[]{\includegraphics[width=0.38\textwidth]{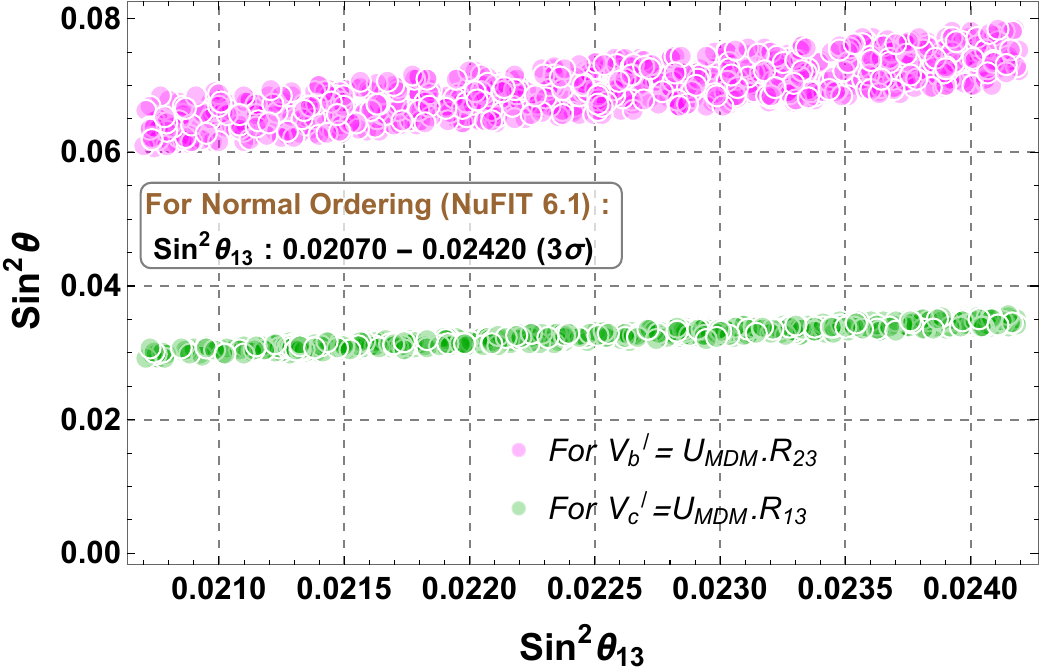}\label{fig:d}} 
    \caption{The correlation plots between (a) $\sin^2\theta_{23}$ and $\cos \alpha$, (b) $\sin^2\theta_{12}$ and $\epsilon$, (c) $\theta$ and $\epsilon$, and (d) $\sin^2\theta_{13}$ and $\sin^2\theta$ are presented for normal ordering. }
\label{fig:1}
\end{figure*}
\begin{figure*}
  \centering
    \subfigure[]{\includegraphics[width=0.38\textwidth]{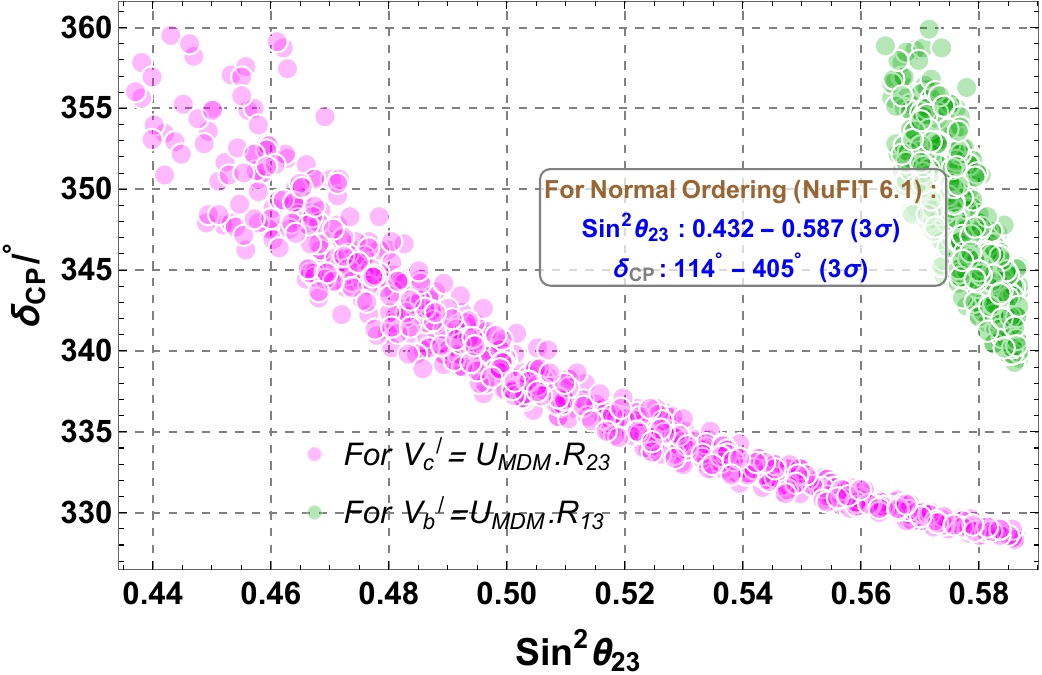}\label{fig:a}} 
    \subfigure[]{\includegraphics[width=0.38\textwidth]{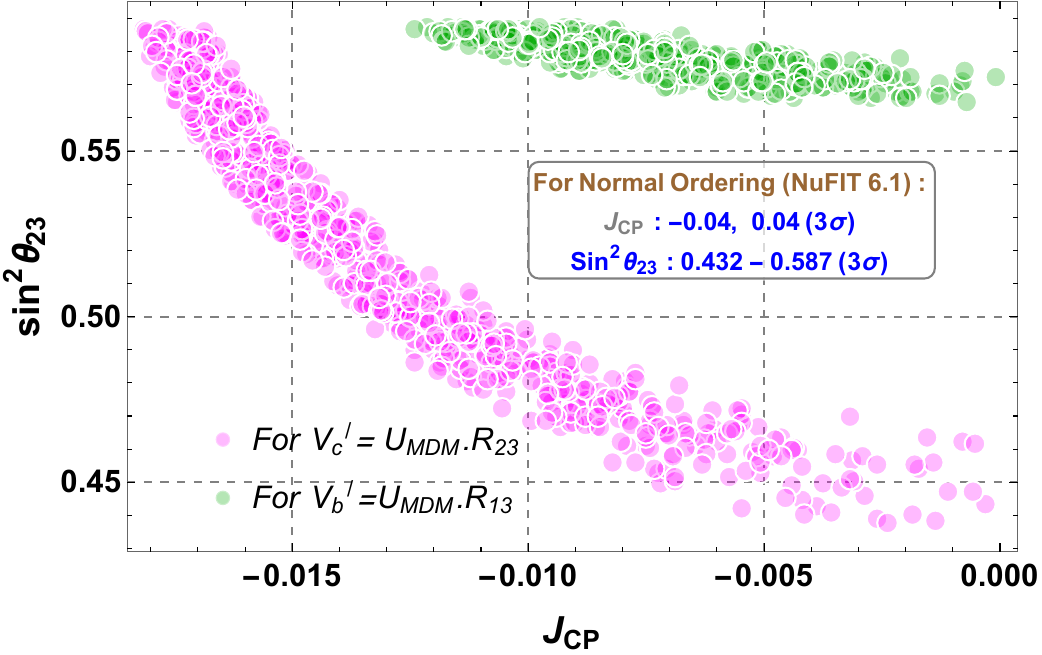}\label{fig:b}} 
    \subfigure[]{\includegraphics[width=0.38\textwidth]{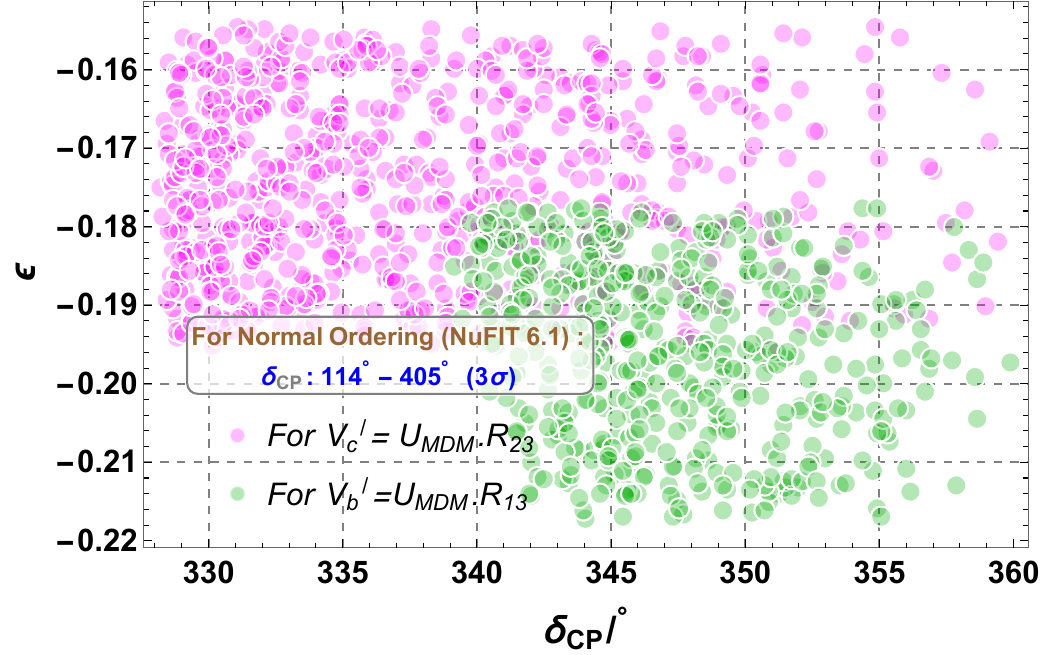}\label{fig:c}} 
    \subfigure[]{\includegraphics[width=0.38\textwidth]{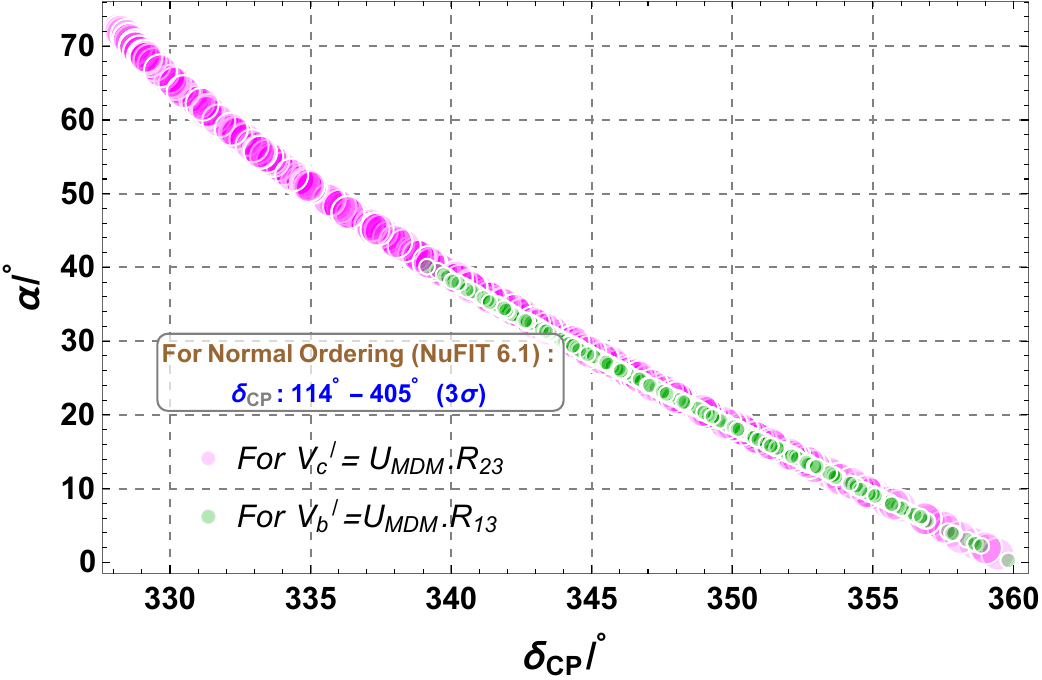}\label{fig:d}} 
    \caption{The correlation plots between (a) $\sin^2\theta_{23}$ and $\delta_{CP}$, (b) $J_{CP} $ and $\sin^2 \theta_{23}$, (c) $\delta_{CP}$ and $\epsilon$, and (d) $\delta_{CP}$ and $\alpha$ are presented for normal ordering. }
\label{fig:1}
\end{figure*}
\subsection{Rotation in the (2-3) sector of the perturbed democratic structure: Second Form}
Now we apply a complex rotation in the (2,3) sector as 
\begin{equation}
    V^{\prime}_c=U_{MDM}.R_{23}
    \label{V1}
\end{equation}
\begin{equation}
    \left(
\begin{array}{ccc}
 A & B c & B e^{i \alpha } s \\
 -G & c F-\frac{e^{-i \alpha } s}{\sqrt{2}} & \frac{c}{\sqrt{2}}+e^{i \alpha } F s \\
 -G & c F+\frac{e^{-i \alpha } s}{\sqrt{2}} & e^{i \alpha } F s-\frac{c}{\sqrt{2}} \\
\end{array}
\right)
\end{equation}
where $A=\sqrt{\frac{2}{3}-\epsilon } , \; B= \sqrt{\epsilon +\frac{1}{3}}, \; F= \sqrt{\frac{1}{3}-\frac{\epsilon }{2}}, G= \sqrt{\frac{\epsilon }{2}+\frac{1}{6}}, \; c= \cos \theta \; \text{and} \; s= \sin \theta
$
\\
\\ 
Comparing Eqn. \ref{V1} with the $U_{PMNS}$ matrix we can evaluate the mixing angles as
\begin{eqnarray}
    \sin^2 \theta_{12} &=& \frac{\left(\epsilon +\frac{1}{2}\right) - \sin ^2\theta_{13} }{1-  \sin ^2\theta_{13}} \label{s12:sec}\\
     \sin^2 \theta_{13} &=& \left(\epsilon +\frac{1}{2}\right) \sin^2\theta \label{s13:sec}\\
      \sin^2 \theta_{23} &=& {1\over 1-s_{13}^2}
\bigg[\left(\frac16-\frac{\epsilon}{3}\right)\sin^2\theta
+\frac23\cos^2\theta \nonumber\\
& & -\frac23\sqrt{1-2\epsilon}\,
\sin\theta\cos\theta\cos\alpha
\bigg]
\label{s23:sec}
\end{eqnarray}

Compairing Eqn. \ref{Jarlskog} with the matrix elements of $V^{\prime}_c$, we get
\begin{equation}
\boxed{
J_{\rm CP}
=
-\frac{(1+2\epsilon)\sqrt{1-2\epsilon}}
{6\sqrt3}
\sin\theta\cos\theta\sin\alpha
}.
\label{eqn:J2}
\end{equation}

We take benchmark values from the allowed $3\sigma$ region in NuFIT 6.1 as
\begin{equation}
s_{12}^2\simeq0.304,\qquad
s_{13}^2\simeq0.0223,
\end{equation}
this yields
\begin{equation}
\epsilon\simeq-0.1804,
\qquad
\sin^2\theta\simeq \; 0.0697,
\qquad
\theta\simeq \; 15.48^\circ .
\end{equation}

We perform parameter scan  for the free parameters $\epsilon$, $\alpha$ and $\theta$ as shown in Table 1 to accommodate the mixing angles within the NuFIT 6.1 constraint. This gives us the Jarlskog Invariant as
\begin{equation}
J_{\rm CP}\simeq-0.0183\sin\alpha,
\end{equation}
so 
\begin{equation}
\boxed{-0.0168\lesssim J_{\rm CP}\lesssim 0.0168}.
\end{equation}
for $ -0.9240\lesssim \sin \alpha\lesssim 0.9240$.
\\
Hence, the Jarslog Invariant deduced from the second form is compatible with the current experimental bound,  $J_{CP} \; \in \; (-0.04, \; 0.04)$.
\\

We scrutinize the solar mixing angle for the different textures analyzed so far by taking the reactor angle as input value from the NuFIT 6.1 and summarize the viability of the texture as follows

\begin{center}
\begin{tabular}{lc}
\toprule
Texture & $\theta_{12}$ \\
\hline
\\
\midrule
$V_a=U_{DM}R_{12}$ & Excluded  \\
$V_b=U_{DM}R_{13}$ & Excluded\\
$V_c=U_{DM}R_{23}$ & Excluded \\
\\
$V^{\prime}_a=U_{MDM}R_{12}$ &  Excluded \\
$V^{\prime}_b=U_{MDM}R_{13}$ & Compatible \\
$V^{\prime}_c=U_{MDM}R_{23}$ & Compatible \\
\bottomrule
\end{tabular}
\end{center}

Thus it is seen that the solar mixing angle can be reproduced within the recent JUNO limit and NuFIT 6.1 for only the textures $V^{\prime}_b$ and $V^{\prime}_c$. We will conduct numerical analysis for only these two viable textures and present their results in the next section.
\section{Result}
 \begin{table}
\centering
\begin{tabular}{ ccc}
\hline
\hline
 \multicolumn{3}{|c|}{\small{Normal Ordering }}
    \\ \hline \hline
\small{Input } & \small{$1^{st}$ Form, $V^{\prime}_b$} &  \small{$2^{nd}$ Form, $V^{\prime}_c$} \\
\small{Parameters} & ($3\sigma$) & ($3\sigma$) \\
\hline
\hline
\small{$\epsilon$} & \small{ -0.2174 $\rightarrow$ -0.1776} &  \small{-0.1956 $\rightarrow$ -0.1547} \\
\hline
\small{$\theta/^{\circ}$} & \small{9.80 $\rightarrow$  10.85} & \small{14.2 $\rightarrow$ 16.2}\\
\hline
\small{$\cos \alpha /{rad}$} & \small{0.765 $\rightarrow$ 0.999} & \small{0.3048 $\rightarrow$ , 0.9998}\\
\hline
\end{tabular}
\caption{Shows the $3\sigma$ ranges of the input parameters $\sin^2\theta_{13}$, $\epsilon$ and $\theta$ taken to find the mixing angles, $\sin^2\theta_{12}$, $\sin^2\theta_{23}$ and $\cos \alpha$. The value of $\sin^2\theta_{13}$ is taken from NuFIT 6.1 for Normal Ordering and the rest three are varied to fit the solar and atmospheric mixing angles within their NuFIT 6.1 bounds.  }
\label{tab:1}
\end{table}
 \begin{table}
\centering
\begin{tabular}{ ccc}
\hline
\hline
 \multicolumn{3}{|c|}{\small{Normal Ordering }}
    \\ \hline \hline
\small{Derived } & \small{$1^{st}$ Form, $V^{\prime}_b$} &  \small{$2^{nd}$ Form, $V^{\prime}_c$} \\
\small{Parameters} & ($3\sigma$) & ($3\sigma$) \\
\hline
\hline
\small{$\sin^2\theta_{12}$}  & \small{0.2893 $\rightarrow$ 0.3294}  & \small{0.2894 $\rightarrow$  0.3294}\\
\hline
\small{$\sin^2\theta_{13}$}  & \small{0.02070 $\rightarrow$ 0.02420}  & \small{0.02070 $\rightarrow$ 0.02420}\\
\hline
\small{$\sin^2\theta_{23}$} & \small{ 0.5642 $\rightarrow$ 0.5869} &  \small{0.4373 $\rightarrow$ 0.5864} \\
\hline
\hline
\small{$J_{CP}$} & \small{ -0.0124 $\rightarrow$ -0.00006} &  \small{-0.0181 $\rightarrow$ -0.0003} \\
\hline
\small{$\delta_{CP}/^\circ$} & \small{ 339.18 $\rightarrow$ 359.88} &  \small{328.24 $\rightarrow$ 359.48} \\
\hline
\end{tabular}
\caption{Shows the range of the solar and atmospheric mixing angles $\sin^2\theta_{12}$ and $\sin^2\theta_{23}$ corresponding to the input parameters in $3\sigma$ ranges as shown in Table 1. The NuFIT 6.1   $3\sigma$ ranges are $\sin^2\theta_{12} \in (0.2893, \; 0.3295)$ and $\sin^2\theta_{23} \in (0.432, \; 0.587)$ respectively.}
\label{tab:2}
\end{table}
For the rotational perturbation in the modified democratic scenarios, we  take the parameters $\epsilon, \; \theta $ and $\cos \alpha$ as input parameters as do parameter scan for a range as shown in Table 1.  We use Eqns \ref{s12:first}, \ref{eqn:theta} and \ref{eqn:theta23} for the first form and Eqns \ref{s12:sec}, \ref{s13:sec} and \ref{s23:sec} for the second form to derive the solar, reactor and atmospheric mixing angles respectively. Their corresponding ranges are presented in Table 2. In Figs 1(a), 1(b), 1(c) and 1(b) we present the correlation plots between (a) $\sin^2\theta_{23}$ and $\cos \alpha$, (b) $\sin^2\theta_{12}$ and $\epsilon$, (c) $\theta$ and $\epsilon$, and (d) $\sin^2\theta_{13}$ and $\sin^2\theta$ are presented for normal ordering.\\
\\
In Fig 1(a), the plot shows a good correlation between the atmospheric angle $\sin^2\theta_{23}$ and the phase parameter $\cos \alpha$ . For a given input range of $  \cos\alpha \in (0.765, 0.999)$, the derived atmospheric mixing angle has a  range of $ \sin^2 \theta_{23} \in (0.5642, 0.5869)$, for the texture $V^{\prime}_{b}$. Similarly, for texture $V^{\prime}_{c}$, the atmospheric mixing angles yield a wider band of $ \sin^2 \theta_{23} \in (0.4373, 0.5864)$, for $  \cos\alpha \in (0.3048, 0.998)$. Both plots confirm that the model-derived ranges of $\sin^2 \theta_{23}$ fall within the allowed range $3\sigma$ $\sin^2 \theta_{23} \in (0.432, \; 0.587)$ as given by NuFIT 6.1. 
\\
\\
Fig 1(b) clearly shows a linear correlation plot between the parameters $\sin^2 \theta_12$ and $\epsilon$ for both the textures $V^{\prime}_{b}$ and $V^{\prime}_{c}$. The solar mixing angles can be reproduced within the allowed NuFIT global data for the textures, by fitting the $\epsilon$ within a range such that it is compatible with Eqns. \ref{s12:first} and \ref{s12:sec} respectively. It is seen that $\epsilon$ chooses a range with higher values for  $V^{\prime}_{c}$ than $V^{\prime}_{b}$. The figure also illustrates that a narrow band of the correlation points falls within the recent JUNO limit, $\sin^2 \theta_{12} \; \in \; (0.3005, 0.3179)$ in the $3\sigma$ range.\\
\\
Fig 1(c) shows a robust evidence that the parameters, $\theta$ and $\epsilon$, are tightly constrained to each other.  Eqns \ref{eqn:theta} and \ref{s13:sec} exhibit the underlying dependence of these two parameters on each other while evaluating the reactor mixing angle, $\sin^2\theta_{13}$. \\
\\
Fig 1(d) highlights the correlation between the reactor mixing angle, $\sin^2 \theta_{13}$ and the rotational angle $\sin^2 \theta$. It can be concluded from the figure that $V^{\prime}_{c}$ demands a larger value of the rotational angle,i.e., $\sin^2 \theta$ than $V^{\prime}_{b}$ such that the reactor angle can be fit within the NuFIT 6.1 allowed range. \\

In Fig 2, we present our analysis corresponding to the charge-parity (CP) violation.  In Fig 2(a), we illustrate the correlation exisiting between the atmospheric mixing angle, $\sin^2 \theta_{23}$ and the CP phase $\delta_{CP}$. The $\delta_{CP}$ can be calculated by comparing Eqns \ref{eqn:J} and \ref{eqn:J1} (or \ref{eqn:J2}) for $V^{\prime}_b$ ($V^{\prime}_c$). The CP phase $\delta_{CP} $ turns out to be highly constrained once the mixing angles are evaluated  from the input parameters, $\alpha$, $\epsilon$ and $\theta$ are fixed. Here, $V^{\prime}_c$ exhibits a broader interval of $\delta_{CP}$ than $V^{\prime}_b$ .\\

Fig 2(b) clearly shows that for the model-derived interval of $\sin^2 \theta_{23}$, i.e., $ 0.5642 \leq \sin^2 \theta_{23} \leq 0.5869 $ for $V^{\prime}_b$ ($ 0.4373 \leq \sin^2 \theta_{23} \leq 0.5864 $ for $V^{\prime}_c$), $J_{CP}$ exhibits a numerical range of $-0.0124 \leq \; J_{CP} \leq -0.00006$ for $V^{\prime}_b$ ($-0.0181 \leq \; J_{CP} \leq -0.0003$ for $V^{\prime}_c$) which falls within the current experimental limit $J_{CP} \in (-0.04, 0.04)$. \\

Fig (2c) and Fig (2d) demonstrates how the model predicted values of the Dirac CP phase $\delta_{CP}$ are affected by the perturbation parameter $\epsilon$ and the rotational phase $\alpha$ respectively. The $\delta_{CP}$ exhibits a scattered plot with $\epsilon$, while the $\delta_{CP}$ shows behaves as a  monotonically decreasing function of $\alpha$ in Fig 2(c) and Fig 2(d) respectively. Thus, $\delta_{CP}$ is not allowed any arbitrary range, as it is highly restricted by the two input parameters $\epsilon$ and $\alpha$.  
\section{Conclusion}
In this work, we have revisited the \enquote{democratic} mixing ansatz which was once ruled out by the experimental findings of neutrino. With an attempt to reconcile its structure and make it compatible with the recent solar angle measurement reported by JUNO experiment and NUFIT 6.1, we first implement rotational corrections along (1-2), (1-3) and (2-3) sectors in the unperturbed democratic mixing matrix, $U_{DM}$ as $V_k =U_{DM}.R_{ij} \;  (i,j,k =1,2,3 \; \text{provided}\; i=j \; \& \; i<j)$  and examine their viability with the current experimental constraints of neutrino parameters.  $V_a =U_{DM}.R_{12}$ is disfavored immediately, as it predicts $U_{e3}=0$, which is not compatible with the current experimental measurement of the reactor mixing angle, $\sin^2 \theta_{13}$. The rotational correction in the (1-3) sector,  $V_b =U_{DM}.R_{13}$, predicts a large solar mixing angle $\sin^2\theta_{12}$ which is far from the 3$\sigma$ interval as reported by JUNO, $\sin^2\theta_{12} = 0.309 \pm 0.0087$. Hence, this texture is excluded too. In  $V_c =U_{DM}.R_{23}$ correction, if we choose a benchmark value of $\sin^2\theta_{13}$ from the allowed range of NuFIT 6.1, say, $\sin^2\theta_{23} \approx \; 0.0222$ then from Eqn. \ref{eqn8-9}, we get $\sin^2 \theta_{12} \approx 0.4886$ which is far from the 3$\sigma$ range of $\sin^2 \theta_{12}$ in NuFIT 6.1. Thus, this texture is disallowed as well.  \\

To fix this limitation in reproducing mixing angles within the allowed constraints of NuFIT 6.1, we introduce a real perturbation parameter $\epsilon$ in $U_{DM}$ such that the unitarity of the matrix is conserved. We denote this new \enquote{modified democratic mixing} as $U_{MDM}$. We implement rotational correction in $U_{MDM}$ along (1-2), (1-3) and (2-3) sectors. This gives us only two viable textures $V^{\prime}_b=U_{MDM}.R_{13}$ and $V^{\prime}_c=U_{MDM}.R_{23}$, while $V^{\prime}_a==U_{MDM}.R_{12}$ is rejected as it gives $U_{e3}=0$. For $V^{\prime}_b$ and $V^{\prime}_c$, the parameters $\epsilon, \; \theta$ and $\alpha$ are scanned in a chosen range such that the mixing angles and $\delta_{CP}$ phase can be accommodated numerically within the allowed interval of NUFIT 6.1. Here we have compared the model predicted values with the experimental values of NuFIT 6.1 for only normal ordering. One can do a similar exercise for inverted ordering too.\\

An important result from this analysis highlights that the atmospheric mixing angle $\sin^2 \theta_{23}$  predicted by $V^{\prime}_c$  ($0.4373 \leq \sin^2 \theta_{23} \leq 0.5864$) has a broader range than $V^{\prime}_b$ ($0.5642 \leq \sin^2 \theta_{23} \leq 0.5869$). Also, the Dirac CP phase $\delta_{CP}$ predicted by  $V^{\prime}_c$ contains a larger allowed region  ($328.24^\circ \leq \delta_{CP} \leq 359.48^\circ$) than $V^{\prime}_b$ ($339.18^\circ \leq \delta_{CP} \leq 359.88^\circ$).  The ranges of the input parameters within which the parameter scan has been performed are shown in Table 1. The numerically obtained ranges of the unknown parameters $\sin^2\theta_{12}, \; \sin^2\theta_{13} $, $\sin^2\theta_{23}$ and $\delta_{CP}$ are shown in Table 2. Fig 1 shows how tightly constrained the mixing angles are against the free parameters $\epsilon$, $\theta$ and $\alpha$.\\

Apart from the above mentioned numerical analysis, we have also investigated Jarlskog Invariant in these two allowed mixing structures.  The predicted values of $J_{CP}$ in $V^{\prime}_b$ is $ J_{CP} \in (-0.0124, -0.00006)$ and that for $V^{\prime}_c$ is $ J_{CP} \in (-0.0181, -0.0003)$, which falls within the current experimental range $J_{CP} \in (-0.04, 0.04)$. Fig 2 gives a clear indication that $J_{CP}$ is strongly  constrained by the parameters the mixing angles ($sin^2\theta_{12},\; sin^2\theta_{13},\;  sin^2\theta_{23}$), $\delta_{CP}$ phase and by the input parameters $\epsilon, \; \theta $ and $\alpha$.   
 \\
Hence, the original democratic mixing ansatz can be made phenomenologically viable by incorporating real perturbation terms in its elements while subjecting it to rotation in the sectors  (1-3) and (2-3). We also conclude that the texture $V^{\prime}_c$ predicts broader allowed regions of the atmospheric mixing angle, the Dirac CP phase and the Jarlskog Invariant than $V^{\prime}_b$ indicating that $V^{\prime}_c$ is more phenomenologically flexible and predictable than $V^{\prime}_b$.  Thus, our rotational perturbation correction to the democratic mixing ansatz provides a predictive framework without ruining the underlying structure of $U_{DM}$. Future experimental measurements will further tell us whether these two viable textures would pass the test of predictability or not.
\bibliographystyle{iopart-num}
\bibliography{ref.bib}
\end{document}